\documentclass[letterpaper,aps,prl,twocolumn,showpacs,groupedaddress]{revtex4}

\usepackage{graphicx}
\usepackage{amsmath}

\begin{document}


\title{Pure spin current from one-photon absorption of linearly polarized light in noncentrosymmetric
semiconductors}


\author{R.D.R Bhat}
\author{F. Nastos}
\author{Ali Najmaie}
\author{J.E. Sipe}
\affiliation{Department of Physics, University of Toronto, 60 St.
George Street, Toronto, Ontario, Canada M5S 1A7}


\date{\today}

\begin{abstract}
We show that one-photon absorption of linearly polarized light
should produce pure spin currents in noncentrosymmetric
semiconductors, including even bulk GaAs. We present $14 \times 14$
$\mathbf{k}\cdot \mathbf{p} $ model calculations of the effect in
GaAs, including strain, and pseudopotential calculations of the
effect in wurtzite CdSe.
\end{abstract}

\pacs{72.25.Fe, 72.25.Dc, 72.40.+w}

\maketitle


There has been a growing interest in the manipulation of electron
spin in semiconductors \cite{wolfSPINSAKPhysTodayZutic,
AwschalomBook, Optical Orientation}. Electron spin provides a new
degree of freedom to be utilized in novel devices; indeed, several
semiconductor `spintronics' devices have been proposed
\cite{DD_DasSarma_Rudolph}. Some of these require a spin-polarized
electrical current (SPEC) \cite{AwschalomBook}. More exotic is a
pure spin current (PSC), in which there is no net motion of charge;
spin up electrons travel in one direction while spin down electrons
travel in the opposite direction. PSCs could result from spin
pumping or Hall effects \cite{PSCs,Malshukov03}, or interference of
two optical beams \cite{BhatSipe,Najmaie}. Some of these have been
observed \cite{StevensPRL,HubnerPRL,Watson03etal}.

In this Letter, we show theoretically that a ballistic PSC can be
generated in noncentrosymmetric semiconductors, merely by linear
absorption. This feature of one-photon absorption of linearly
polarized light does not seem to have been appreciated previously:
It occurs even in unstrained bulk GaAs, it can be generated by a
single, weak continuous wave (CW) laser beam, and it arises from the
symmetry of the crystal itself.

We consider a semiconductor with filled valence bands and empty
conduction bands, uniformly illuminated by light with photon energy
larger than the direct band gap. The electric field inside the
semiconductor is $\mathbf{E} \left( t\right) =\mathbf{E}_{\omega
}\exp \left( -i\omega t\right) +c.c.$, where $\mathbf{E}_{\omega }$
is a slowly varying envelope function. The coupling between the
light and electrons is treated in the long-wavelength limit, and we
ignore interactions amongst the electrons. This is a common starting
point for calculations of interband absorption in semiconductors
\cite{YuCardona}, and we adopt it to show that PSCs can be generated
even within this relatively simple model. We use the semiconductor
optical Bloch equations (SOBEs) solved perturbatively to first order
in the field intensity \cite{HaugKochbookRossler02}. The effect we
consider does not rely on carrier scattering, but to connect with
experiments, we include an estimate of the spin separation using a
simple model of momentum relaxation.

We indicate below that, within these approximations, one-photon
absorption of linearly polarized light excites a distribution of
electrons in $\mathbf{k}$-space that is even in $\mathbf{k}$, no
matter what the symmetry of the material; no net electrical
current results \cite{footLPGE}. Further, electrons excited into
the conduction band at opposite $\mathbf{k}$ points will have
opposite spin polarization, resulting in no net spin injection. In
noncentrosymmetric crystals however, the spin polarization
injected at a given $\mathbf{k}$ need not vanish
\cite{footAlvarado}. Thus, in such a case, there will be a PSC,
since the velocities of electrons at opposite $\mathbf{k}$ points
are opposite.

By treating the field perturbatively, and assuming fast interband
dephasing, the SOBE gives for the single particle density matrix
$\rho_{nm}\left( t\right)$,
\begin{equation}
\begin{split}
\rho _{nm} =& \rho _{nm}^{(\text{ini})}+
\sum_{v}\int_{-\infty}^{t}dt^{\prime } \frac{\mathbf{E}_{\omega
}\left( t^{\prime }\right)\cdot \mathbf{v}_{nv}}{\omega_{nv}}
\frac{\mathbf{E}_{\omega }^{*}\left( t^{\prime }\right)
 \cdot \mathbf{v}_{vm}}{\omega _{mv}}\\
& \times  \frac{\pi e^{2}}{\hbar ^{2}  } e^{ \left( -i\omega
_{nm}-\gamma _{nm}\right) \left( t-t^{\prime }\right) } \left(
D_{nv} +D^{*}_{mv}\right),\label{e:SOBEsoln}
\end{split}
\end{equation}
where $\rho _{nm}^{(\text{ini})}$ is the initial density matrix,
equal to $1$ when $n=m$ is a valence band and equal to $0$
otherwise, $\omega _{nm}\left( \mathbf{k}\right) \equiv \omega
_{n}\left( \mathbf{k}\right) -\omega _{m}\left( \mathbf{k}\right) $,
$\hbar \omega _{n}\left( \mathbf{k}\right) $ is the energy of the
Bloch state $\left| n\mathbf{k} \right\rangle $, the velocity matrix
element $\mathbf{v}_{cv}\left( \mathbf{k}\right) \equiv \left\langle
c \mathbf{k}\right| \mathbf{\hat{v}}\left| v\mathbf{k}\right\rangle
$, $\gamma_{nm}\left(\mathbf{k}\right)$ is the dephasing rate, and
$D_{nv} \equiv \delta \left( \omega -\omega _{nv}\right) -\left(i/
\pi \right)\mathcal{P} \left(\omega -\omega _{nv}\right)^{-1}$,
where $\mathcal{P}$ indicates the principal part. The dephasing
rates approximate many-body interactions
\cite{HaugKochbookRossler02}.

The photoexcited electron density is $N=\sum_{c,\mathbf{k}}\rho
_{cc,\mathbf{k}}\left( t\right)$. Defining $N_{cv}\left(
\mathbf{k}\right)$ as the electron density excited from band $v$
to band $c$ at $\mathbf{k}$ (i.e.
$N=\sum_{\mathbf{k},c,v}N_{cv}\left( \mathbf{k}\right) $), we find
from Eq.\ (\ref{e:SOBEsoln}) that $\left(
d/dt+\gamma_{cc}\right)N_{cv}\left( \mathbf{k}\right)
=\dot{N}^{(\text{inj})}_{cv}\left(\mathbf{k}\right),$ where
\[
\dot{N}^{(\text{inj})}_{cv}\left(\mathbf{k}\right) \equiv
\frac{2\pi e^{2}}{\hbar ^{2}\omega ^{2} }\frac{1}{V}\left|
\mathbf{E}_{\omega }\cdot \mathbf{v}_{cv}\left( \mathbf{k} \right)
\right| ^{2}\delta \left( \omega _{cv}\left( \mathbf{k}\right)
-\omega \right),
\]
and $V$ is a normalization volume \cite{YuCardona}. Time reversal
symmetry of the Bloch states results in Kramers degeneracy, $\omega
_{n}\left( -\mathbf{k}\right) =\omega _{\bar{n}}\left(
\mathbf{k}\right) $, where a bar above a band index denotes the band
with the opposite spin. As well, the velocity matrix elements
satisfy $\mathbf{v} _{cv}\left( -\mathbf{k}\right) =-\left[
\mathbf{v}_{\bar{c}\bar{v}}\left( \mathbf{k}\right) \right] ^{*}$
\cite{MatrixElementSymmetry}. Using these two properties, it follows
that when the light is linearly polarized,
$\dot{N}^{(\text{inj})}_{cv}\left(\mathbf{-k}\right)
=\dot{N}^{(\text{inj})}_{\bar{c}\bar{v}}\left(\mathbf{k}\right)$.
Consequently, the photocurrent injection rate, given by
$e\sum_{\mathbf{k},c,v}\left[ \mathbf{v}_{cc}\left(
\mathbf{k}\right) -\mathbf{v}_{vv}\left( \mathbf{k}\right)
\right]\dot{N}^{(\text{inj})}_{cv}\left(\mathbf{k}\right)$, is zero
for linearly polarized light; hence, any spin current must be a
PSC\@.

The photoexcited electron spin density is
$\mathbf{S}=\sum_{c,c^{\prime },\mathbf{k}}\left\langle c^{\prime
}\mathbf{k}\right| \mathbf{\hat{S}}\left| c \mathbf{k}\right\rangle
\rho _{cc^{\prime },\mathbf{k}}\left( t\right)$, where
$\mathbf{\hat{S}}$ is the spin operator. We neglect any spin
polarization of the holes, since their spin relaxation times are
typically very short \cite{HiltonTang}. Note that for excitation
near the band edge $\hbar \omega_{cc^{\prime }}$ is either less than
a few meV (if $c$ and $c^{\prime}$ are equal, or are the spin-split
lowest conduction bands) or greater than a few eV (if $c$ or
$c^{\prime}$ is a higher conduction band) \cite{Dresselhaus_CCF88}.
At times $t$ longer than the pulse width $t_{L}$, the integral over
$t^{\prime}$ in Eq.\ (\ref{e:SOBEsoln}) will be negligible unless
$\omega_{cc^{\prime }}<1/t_{L}$. For long pulses, a similar argument
can be made but with $\gamma_{cc^{\prime}}$ replacing $1/t_{L}$;
however, in this Letter, we focus on typical ultrafast experiments
\cite{StevensPRL,HubnerPRL}, for which one can neglect spin
relaxation and carrier recombination occurring on longer timescales.
Thus the integral over $t^{\prime}$ allows one to neglect coherences
other than those between spin-split pairs of bands. We expand the
retained coherences in powers of $\omega_{cc^{\prime }}/\omega$ and
keep only the lowest order term. We thus neglect the precession of
the spins due to the spin-splitting of the bands; this is justified
since the precession period is long compared to the momentum
scattering time \cite{Optical Orientation}. Writing
$\mathbf{S}=\sum_{\mathbf{k}}\mathbf{S}\left( \mathbf{k}\right) $,
where $\mathbf{S}\left( \mathbf{k}\right) $ is the electron spin
density at $\mathbf{k}$ we find $S^{i}\left(
\mathbf{k}\right)=\sum_{j,l}\zeta^{ijl}(\mathbf{k})
\int_{-\infty}^{t}E_{\omega }^{j*}\left( t^{\prime }\right)
E_{\omega }^{l}\left( t^{\prime }\right) dt^{\prime }$, where
superscript indices denote Cartesian components,
\begin{equation}
\begin{split}
\zeta^{ijl}(\mathbf{k}) \equiv & \frac{\pi e^{2}}{\hbar ^{2} \omega
^{2}}\frac{1}{V}\sum_{v,c,c^{\prime }}^{\prime} \left\langle
c^{\prime }\mathbf{k}\right| \hat{S}^{i} \left| c
\mathbf{k}\right\rangle v^{j}_{vc^{\prime }}\left( \mathbf{k}\right)
v^{l} _{cv}\left(
\mathbf{k}\right)\label{Sdot_w_spinsplitting}\\
&\times \left[ \delta \left( \omega _{cv}\left( \mathbf{k}\right)
-\omega \right) +\delta \left( \omega _{c^{\prime }v}\left(
\mathbf{k}\right) -\omega \right) \right],
\end{split}
\end{equation}
and the prime on the summation indicates a restriction to pairs
$\left( c,c^{\prime} \right)$ for which either $c^{\prime}=c$, or
$c$ and $c^{\prime }$ are a quasi-degenerate pair. Thus the
coherence between bands $c$ and $\bar{c}$ is optically excited and
grows with the population, as in simpler band models that neglect
spin splitting \cite{Optical Orientation}. If one were to neglect
the coherence between spin-split bands, the net degree of spin
polarization, $|\mathbf{S}|/\left(N\hbar /2\right)$, would only be
$10\%$ in GaAs compared to the accepted $50\%$ with Eq.\
(\ref{Sdot_w_spinsplitting}).

From the time reversal properties of the Bloch states, the spin
matrix elements satisfy $\left\langle n,-\mathbf{k}\right|
\mathbf{\hat{S}}\left| m,-\mathbf{k}\right\rangle =-\left\langle
\bar{n},\mathbf{k}\right| \mathbf{ \hat{S}}\left|
\bar{m},\mathbf{k}\right\rangle ^{*}$ \cite{MatrixElementSymmetry}.
Using this property, Kramers degeneracy, and the time reversal
property of the velocity matrix elements, one finds
$\mathbf{S}\left( -\mathbf{k} \right) =-\mathbf{S}\left(
\mathbf{k}\right) $ for linear polarized excitation. Thus there is
no net spin injection from linearly polarized light, and if
$\mathbf{S}\left( \mathbf{k} \right)$ is nonzero for some
$\mathbf{k}$, there is a PSC\@.

A PSC can be quantified by $\left( I_{\uparrow }-I_{\downarrow
}\right) $, where $I_{\uparrow \left( \downarrow \right) }$ is the
current due to up (down) electrons. A more general measure that
naturally accounts for a distribution of carrier velocities and
spins is to use a spin current density pseudotensor $K^{ij}\equiv
\left\langle \hat{v}^{i}\hat{S}^{j}\right\rangle $
\cite{BhatSipe}.

Phenomenologically, the injection of a spin current due to
one-photon absorption can be written in terms of a material response
pseudotensor $\mu $, as $K^{ij}=\sum_{l,m}\mu^{ijlm} \int E_{\omega
}^{l*}\left( t^{\prime }\right) E_{\omega }^{m}\left( t^{\prime
}\right) dt^{\prime }$. The spin current pseudotensor $\mu $
satisfies the intrinsic symmetry $\mu ^{ijlm}=\left(
\mu^{ijml}\right) ^{*}$. It is further constrained by the symmetry
of the material, since it must be invariant under the point group
transformations of the crystal. It vanishes for materials with
inversion symmetry, but can be nonzero for the symmetries
appropriate to zincblende and wurtzite crystals.

The microscopic expression for $\mu$ is derived from Eq.\
(\ref{e:SOBEsoln}), with the same approximations made for Eq.\
(\ref{Sdot_w_spinsplitting}). Using time reversal properties of the
Bloch functions, one can then show that $\mu $ is real, and can be
written as
\begin{equation}
\begin{split}
\mu ^{ijlm} =&\frac{\pi e^{2}}{\hbar ^{2}\omega
^{2}}\frac{1}{V}\sum_{ \mathbf{k},v}\sum_{c,c^{\prime }}^{\prime
}\delta \left( \omega _{cv}\left(
\mathbf{k}\right) -\omega \right)  \label{e:mu}\\
&\times \mathrm{Re} \left[ \left\langle c^{\prime
}\mathbf{k}\right| \hat{v} ^{i}\hat{S}^{j}\left|
c\mathbf{k}\right\rangle v_{vc^{\prime }}^{l}\left(
\mathbf{k}\right) v_{cv}^{m}\left( \mathbf{k}\right) +\left(
l\leftrightarrow m\right) \right] .
\end{split}
\end{equation}

In an ultrafast experiment, a PSC can be measured by its
displacement of up and down spins. To estimate the spin separation,
we use the optically injected electron distribution as a source term
in the Boltzmann transport equation. By neglecting space-charge
effects, which are negligible for a PSC, the Boltzmann equation can
be solved in the relaxation time approximation \cite{HubnerPRL}.
This approach neglects scattering during optical excitation, but for
ultrafast experiments the error should not be too large. If one
measures the spin with respect to the quantization direction
$\mathbf{\hat{a}}$, the up and down spin populations are separated
by an average displacement $\mathbf{d}\left( \mathbf{\hat{a}}
\right) $. We find $d^{i}\left( \mathbf{\hat{a}}\right)
=\sum_{j}\left( 4\tau /\hbar \right) K^{ij}\hat{a}^{j}/N$, where
$\tau $ is the momentum relaxation time.

To calculate the PSC for bulk GaAs, we use a $\mathbf{k}\cdot
\mathbf{p} $ model that diagonalizes the one-electron Hamiltonian
(including spin-orbit coupling) within a basis set of 14 $\Gamma $
point states and includes important remote band effects \cite{PZ96}.
The 14 states (counting one for each spin) comprise six valence band
states (the split-off, heavy, and light hole bands), and eight
conduction band states (the two lowest, which are $s$-like, and the
six next-lowest, which are $p$-like). The model contains 13
parameters chosen to fit low-temperature experimental data
\cite{PZ96}.

Strain is included by the deformation potential method of Pikus and
Bir \cite{PikusBirBirPikus}. The deformation potentials amongst
valence and lowest conduction band states are well established
\cite{Vurgaftman01}. Between the valence and $p$-like conduction
band states, there are two deformation potentials for the strain we
consider, $a_{cv}$ and $b_{cv}$ \cite{Koopmans98, Bertho94,
Blacha84}. We use $b_{cv}=-2.3$\,eV, which is an average of tight
binding calculations \cite{Koopmans98, Bertho94}, and consistent
with experiment \cite{Etchegoin92}. The parameter $a_{cv}$ couples
to the hydrostatic component of strain \cite{Bertho94}; neither it
nor the deformation potentials amongst $p$-like upper-conduction
band states affect our results. We also neglect the small effect of
strain on spin-orbit coupling \cite{Bertho94}.

It is interesting to compare the GaAs results with a material of
different symmetry: wurtzite CdSe. Since $\mathbf{k}\cdot \mathbf{p}
$ methods are less developed for wurtzite crystals, we use a
relativistic pseudopotential plane wave code within the local
density approximation \cite{abinitio}; the use of a different method
also illustrates the robustness of the PSC effect. Quasiparticle
effects are included with a scissor correction. The matrix elements
of $\hat{v}^{i}\hat{S}^{j}$ were calculated using $\left\langle
c^{\prime }\mathbf{k}\right| \hat{v}^{i}\hat{S} ^{j}\left|
c\mathbf{k}\right\rangle =\sum_{n}v_{cn}^{i}\left( \mathbf{k}
\right) \left\langle n\mathbf{k}\right| \hat{S}^{j}\left|
c\mathbf{k} \right\rangle $ with 16 valence and 24 conduction bands
included in the sum; nonlocal contributions to these matrix elements
were neglected. The Brillouin zone summation used 2600 irreducible
$\mathbf{k}$ points.

For GaAs, we present results for light linearly polarized along
$\left[ 001\right] $. In the standard cubic basis,
$K^{xx}=-K^{yy}\equiv \kappa$ are the only two nonzero components of
the spin current. Alternately, using a basis where
$\mathbf{\hat{x}}\parallel \left[ 001\right] $,
$\mathbf{\hat{z}}\parallel \left[ 110\right] $ and $\mathbf{
\hat{y}}\parallel \left[ 1\bar{1}0\right] $, the only two nonzero
components of the spin current are $K^{zy}=K^{yz}=\kappa$. That is,
there is a net PSC of electrons with spin component along $\left[
1\bar{1}0\right] $ and velocity component along $\left[ 110\right]
$, and an equal PSC with spin component along $\left[ 110\right] $
and velocity component along $\left[ 1 \bar{1}0\right] $. Under
strain in the $\left[ 001\right] $ direction, the point group
symmetry of the crystal is reduced from $T_{d}$ to $D_{2d}$, but
$K^{zy}=K^{yz}$ remain the only nonzero components of the spin
current. We have considered 1\% biaxial strain; under tensile
(compressive) strain, the lattice dilates (contracts) by 1\% in the
$\left[ 100\right] $ and $\left[ 010\right] $ directions and
contracts (dilates) by 0.93\% in the $\left[ 001\right] $ direction,
with the latter determined from the elastic constants $C_{11}$ and
$C_{12}$ \cite{Vurgaftman01}.

For CdSe, we present results for light linearly polarized at an
angle $\pi /4 $ to the $c$-axis. Defining $\mathbf{\hat{n}}$ as a
vector perpendicular to both the $c$-axis and the light
polarization, we restrict our attention to $\mathbf{d}\left(
\mathbf{\hat{n}}\right) $, which lies in the plane perpendicular
to $\mathbf{\hat{n}}$.

\begin{figure}
\includegraphics{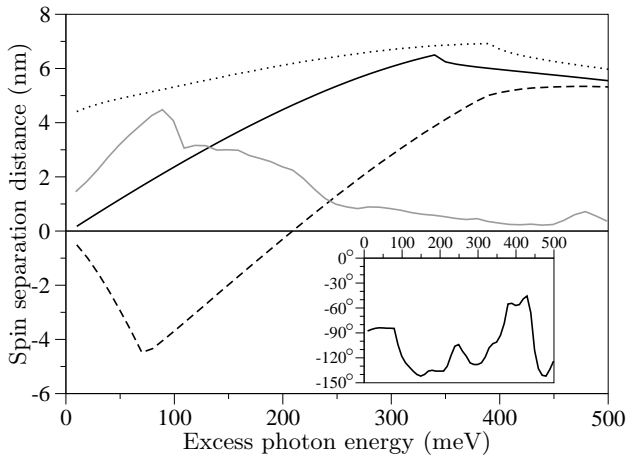}
\caption{\label{spindist} Calculated spin separation in unstrained
GaAs (thick line), in GaAs under 1\% tensile (dotted line) and
compressive (dashed line) strain, and in CdSe (grey line). In GaAs,
transitions from the split-off band begin to contribute at
$\approx350$\,meV. The inset shows, for CdSe, the angle $\theta$
between $\mathbf{d}(\hat{\mathbf{n}})$ and the $c$ axis; $\theta =
45^{\circ}$ corresponds to $\mathbf{d}(\hat{\mathbf{n}}) \parallel
\mathbf{E}_{\omega}$.}
\end{figure}

\begin{figure*}
\includegraphics[width=7.0in]{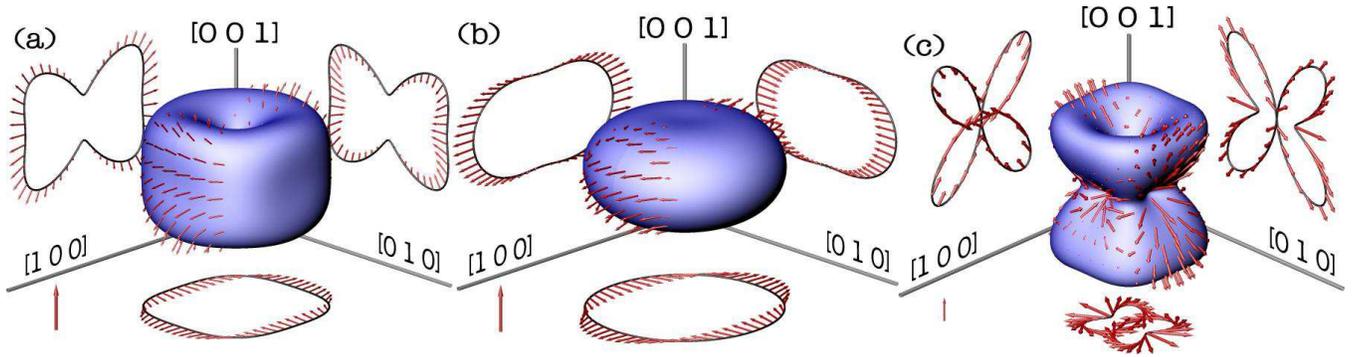}
\caption{\label{distn} (Color online) (a) Calculated
$\mathbf{\hat{k}}$-space distributions of photoexcited electrons
in unstrained GaAs with excess photon energy 300\,meV. The surface
is the injected carrier density, $N( \mathbf{\hat{k}}) $, and the
vectors affixed to the surface give the degree of spin
polarization, $\mathbf{s}( \mathbf{\hat{k}})$, of electrons
excited with crystal momentum in the direction $\mathbf{\hat{k}}$.
The cross sections through the centers of the distributions, along
with the projection of spins on the cross sections, have been
displaced to aid visualization. (b) and (c) are as in (a), but at
20\,meV excess photon energy and for 1\% tensile  and compressive
strains respectively. The reference spin in the lower left corner
of each panel is $0.25 \hbar / 2$.}
\end{figure*}

The calculated spin separation distances, $\left| \mathbf{d}\right|
$, for these PSCs are plotted in Fig.\ 1 as a function of photon
energy above the band gap, where we have assumed a momentum
relaxation time of 100\,fs; note that the band gap for each case is
different. The inset indicates the direction of the PSC in CdSe. For
comparison, experiments on PSCs generated with harmonically related
beams in GaAs/AlGaAs quantum wells \cite{StevensPRL} and ZnSe
\cite{HubnerPRL} measured spin separations of 20\,nm and 24\,nm
respectively.

To illustrate the spin-momentum correlation implied by the
calculation of $\mathbf{d}$, we plot in Fig.\ 2 calculated angular
distributions of the injected carrier density $N( \mathbf{\hat{k}})
\equiv \sum_{\left| \mathbf{k} \right| }N\left( \mathbf{k}\right) $
and degree of spin polarization $\mathbf{s}( \mathbf{\hat{k}})
\equiv \left[ \sum_{\left| \mathbf{k} \right| }\mathbf{S}\left(
\mathbf{k}\right) \right] / N( \mathbf{\hat{k}})$ for the excited
conduction band electrons in GaAs.

In unstrained GaAs, $s_{\textrm{max}}$, the maximum over
$\mathbf{k}$ of $|\mathbf{s}(
\mathbf{\hat{k}})|/\left(\hbar/2\right)$, rises from zero at the
band edge to a maximum of 12\% at the photon energy when transitions
from the split-off valence band become allowed. Under strain,
$s_{\textrm{max}}$ is largest near the band edge. In Fig.\ 2(b),
$s_{\textrm{max}}$ is 15\% while in Fig.\ 2(c) it is 44\%.

It is clear that strain can increase the PSC, especially at lower
energies where strain-induced splitting of the heavy and light
hole bands increases the spin polarization of the photogenerated
electrons. Under tensile (compressive) strain the light hole band
moves to higher (lower) energy than the heavy hole band. The
calculated splitting between heavy and light hole bands at the
$\Gamma $ point is 84\,meV for tensile strain and 69\,meV for
compressive strain. The crystal field splitting between heavy and
light hole bands in CdSe is 38\,meV\@.

Note that in Eq.\ (\ref{e:mu}) we have assumed that
$\mathbf{\hat{v}}$ and $\mathbf{\hat{S}}$ commute. This is no longer
true when the anomalous velocity term, $
\hbar\left(\boldsymbol{\sigma } \times \boldsymbol{\nabla
}V\right)/\left(4m^{2}c^{2}\right)$, is included in
$\mathbf{\hat{v}}$. The PSC we have described here does not require
that term, in contrast to other effects
\cite{Malshukov03}\cite{Rashba03}. In fact, in the $\mathbf{k} \cdot
\mathbf{p}$ calculation, the anomalous velocity is neglected.
Including $k$-dependent spin-orbit coupling between valence and
lowest conduction bands (with $C_{0}=0.16$\,eV {\AA}
\cite{Ostromek96}), and the associated anomalous velocity, changes
the calculated spin separation distances by less than 0.1\,nm.

While the calculations of ballistic PSCs that we have presented
here are appropriate for bulk semiconductors, larger PSCs may be
possible in heterostructures, which can be prepared with large
structural asymmetry. As well, we note that since only linear
absorption is involved, this effect could be studied even with CW
beams, and in nanostructures, where short spin transport distances
could still have significant consequences.

In a material with low enough symmetry, one-photon absorption of
circularly polarized light can generate a ballistic current without
an applied voltage \cite{SturmanFridkin}; this so-called circular
photogalvanic effect (CPGE) is a SPEC
\cite{GanichevReviewBelkovCPGEinterband}. It is tempting to
understand the effect we present here as a superposition of opposite
SPECs due to two fields with opposite helicity. However, the CPGE
cannot occur in crystals with zincblende symmetry, although it can
in wurtzite CdSe \cite{Laman} and strained GaAs
\cite{LyandaGellerPikus}. It is worth noting that in addition to
describing the PSC we present here, the spin current response
pseudotensor in Eq.\ (\ref{e:mu}) can also describe the SPEC due to
the the CPGE\@.

Finally, we point out that a PSC can be generated even with
unpolarized light. Averaging over polarization directions in a
plane, one obtains $K^{ij}=\int E^{2}_{\omega} dt^{\prime}
\sum_{l,m} \left( \delta ^{lm}-\hat{n}^{l}\hat{n}^{m}\right)
\mu^{ijlm}/2$, where $\hat{n}$ is the propagation direction of the
light. This PSC is smaller than that due to linearly polarized
light.

\begin{acknowledgments}
We thank N. Arzate, E. Sherman, Art Smirl and H.M. van Driel for
useful discussions. This work was supported by NSERC, PRO and
DARPA-SpinS.
\end{acknowledgments}


\begin{thebibliography}{99}
\bibitem{footLPGE}  Linear photogalvanic effects, due to asymmetric scattering
or a shift in charge density upon excitation can occur
\cite{SturmanFridkin}, but not within these approximations.
\end{thebibliography}

\end{document}